\pdfoutput=1

\documentclass[10pt]{article}

\usepackage{authblk}                
\usepackage[square,numbers]{natbib} 
\usepackage[T1]{fontenc}
\usepackage[utf8]{inputenc} 

\usepackage{amsmath,amssymb}                

\usepackage{listings}    
\usepackage{graphicx}    
\usepackage{subcaption}  
\usepackage{booktabs}    
\usepackage{hyperref}    
\usepackage{xurl}        
\usepackage[htt]{hyphenat} 
\usepackage{microtype} 
\DisableLigatures{}    

\usepackage{placeins}
\usepackage{setspace}
\usepackage{multirow}
\usepackage{enumerate}
\usepackage{pdflscape}
\usepackage{pgfplots} 

\usepackage{xcolor}
\definecolor{codegreen}{rgb}{0.25,0.5,0.35}
\definecolor{codegray}{rgb}{0.5,0.5,0.5}
\definecolor{codepurple}{rgb}{0.6,0,0}
\definecolor{backcolour}{rgb}{0.95,0.95,0.92}
\definecolor{colorstring}{rgb}{0.5,0,0.35}
\definecolor{rltred}{rgb}{0.5,0,0}
\definecolor{rltgreen}{rgb}{0,0.5,0}
\definecolor{rltblue}{rgb}{0,0,0.5}
\definecolor{DarkGreen}{rgb}{0.00,0.60,0.00}
\definecolor{ScarletRed}{rgb}{0.80,0.00,0.00}
\definecolor{blizzardblue}{rgb}{0.67, 0.9, 0.93}
\definecolor{green-yellow}{rgb}{0.68, 1.0, 0.18}
\definecolor{dkgreen}{rgb}{0,0.6,0}
\definecolor{gray}{rgb}{0.5,0.5,0.5}
\definecolor{mauve}{rgb}{0.58,0,0.82}
\definecolor{lightgrey}{rgb}{0.90,0.90,0.90}
\definecolor{grey}{gray}{0.75}
\definecolor{light-gray}{gray}{0.80}

\lstdefinestyle{mystyle}{
    escapechar=©, 
    language=Python,
	backgroundcolor=\color{backcolour},
    basicstyle=\footnotesize\ttfamily,
   	identifierstyle=\footnotesize\ttfamily,
	commentstyle=\color{codegreen},
	keywordstyle=\color{colorstring}\bfseries,
	morekeywords={OR, AND},
	numberstyle=\ttfamily\color{codegray},
	stringstyle=\ttfamily\color{DarkGreen},
	breakatwhitespace=false,
	breaklines=true,
	captionpos=b,
	keepspaces=true,
	numbers=left, 
	numbersep=2pt,
	showspaces=false,
	showstringspaces=false,
	showtabs=false,
	tabsize=2
}
\usepackage{braket}		
\usepackage{mathtools}	

\usepackage{pifont} 

\usepackage{xspace}

\usepackage{tcolorbox}
\newtcolorbox{resultsbox}[1][]
{
	colframe=gray!100, 
	colback=white!100, 
	coltitle=white,
	title=#1 
}

\usepackage{ifthen}
\newboolean{showcomments}
\setboolean{showcomments}{true} 

\ifthenelse{\boolean{showcomments}}{
	\newcommand{\nbc}[3]{
		{\colorbox{#3}{\bfseries\sffamily\scriptsize\textcolor{white}{#1}}}
		{\textcolor{#3}{\sf\small$\langle$\textit{#2}$\rangle$}}}
	
}{
	\newcommand{\nbc}[3]{}

}

\newcommand{\qready}{\textit{Q-READY}\xspace}

\usepackage{tabularx}
\usepackage{multirow}
\usepackage{booktabs}
\usepackage{ltablex}

\providecommand{\Description}[1]{}

\usepackage{geometry}
\title{
{\qready}: Predictive Feasibility Assessment for Hybrid Quantum-Classical Applications

}

\author[1]{Tao Yue}
\author[1]{Man Zhang}

\affil[1]{Beihang University}

\date{}

\begin{document}

\maketitle

\begin{abstract}
Quantum computing is rapidly evolving into an emerging computational infrastructure and is increasingly being used to tackle real-world problems in domains such as chemistry, materials science, logistics, and finance, as well as software engineering problems such as test optimization and project scheduling. Hybrid quantum-classical applications are particularly important because they provide a practical path for integrating quantum capabilities into existing software systems under near-term hardware constraints. However, the engineering of hybrid quantum-classical applications remains largely ad hoc and constrained by hardware limitations including qubit scarcity, noise, and limited connectivity. In this paper, we propose \qready to address the lack of systematic methodologies for assessing the feasibility of hybrid solutions prior to costly implementation.
Positioned as a Model-Based Systems Engineering (MBSE) approach grounded in Model-Driven Engineering (MDE) principles, \qready establishes a structured pipeline encompassing requirements modeling, problem formulation, workflow design, and hardware-aware feasibility assessment, enabling simulation-based evaluation and comparison of candidate solutions under realistic constraints through traceable system-level models and backend-aware abstractions. We illustrate the pipeline with a running credit-portfolio capital-assessment example, showing how requirements, problem structure, strategy choices, workflow behavior, backend assumptions, and feasibility evidence can be linked into a coherent engineering decision.
\qready is envisioned as an environment that supports executable modeling, constraint evaluation, and predictive analysis. Its expected outcomes include a systematic methodology for hybrid quantum application design, a supporting software platform, benchmark datasets, and empirical design guidelines.

\end{abstract}


\section{Introduction}
\label{sec:Intro}
Quantum computing is transitioning from a predominantly theoretical pursuit to an emerging computational infrastructure that is increasingly accessible and experimentally viable~\cite{aruteQuantumSupremacyUsing2019}. Although current devices remain within the Noisy Intermediate-Scale Quantum (NISQ) regime~\cite{preskillQuantumComputingNISQ2018}, characterized by limited qubit counts and significant noise, quantum processors have progressed from laboratory prototypes to programmable platforms that support practical, multidisciplinary experimentation across domains such as optimization, machine learning, and scientific computing~\cite{leymannBitterTruthQuantum2020}, as well as software engineering~\cite{zhangQuantumOptimizationSoftware2025}.

Cloud-accessible platforms, including IBM Quantum\footnote{See: \url{https://www.ibm.com/quantum}}, Origin Quantum\footnote{See: \url{https://qcloud.originqc.com.cn/en/services\#hybrid}}, Amazon Braket\footnote{See: \url{https://aws.amazon.com/braket/}}, Microsoft Azure Quantum\footnote{See: \url{https://azure.microsoft.com/en-us/solutions/quantum-computing}}, and D-Wave Quantum\footnote{See: \url{https://www.dwavequantum.com/}}, provide programmable access to both gate-based and annealing-based architectures, enabling the development and execution of quantum programs within classical computing environments~\cite{reschQuantumComputingCloud2019,gillQuantumCloudComputing2022}. In parallel, heterogeneous computing architectures that integrate quantum processing units (QPUs) with classical central processing units (CPUs) and graphics processing units (GPUs) increasingly support hybrid quantum--classical workflows, where quantum components act as accelerators within larger computational pipelines~\cite{moguel2022quantum,abbas2021power}. This hybridization is particularly important in the NISQ era because it offers a practical path for embedding limited quantum capabilities into existing software systems while relying on classical components to absorb the remaining workload. These advances, together with growing research and industrial ecosystems such as QuTech\footnote{See: \href{https://qutech.nl/}{https://qutech.nl/}.} and Atos\footnote{See: \href{https://atos.net/}{https://atos.net/}.}, have enabled practical experimentation with hybrid algorithms for optimization, machine learning, and scientific computing~\cite{bhartiNoisyIntermediatescaleQuantum2022,qutechMissiondrivenQuantumResearch2026,atosAtosQuantumLearning2026}.
 
Different definitions of hybrid quantum–classical applications have been proposed in the literature, reflecting varying levels of abstraction. At the algorithmic level, hybrid approaches are typically defined by the tight coupling of parameterized quantum circuits with classical optimization loops, as seen in variational quantum algorithms such as Variational Quantum Eigensolver (VQE) and Quantum Approximate Optimization Algorithm (QAOA), where classical routines iteratively update quantum circuit parameters~\cite{VariationalQuantumAlgorithms2021}. At a broader system level, hybrid applications are viewed as multi-stage computational workflows that integrate classical preprocessing, problem encoding, quantum execution, and classical post-processing, emphasizing end-to-end application structure rather than isolated algorithms~\cite{leymannBitterTruthQuantum2020,wederHybridQuantumApplications2021}. From an infrastructure perspective, hybrid quantum–classical computing refers to the co-execution of classical and quantum resources across heterogeneous platforms, often involving cloud-based orchestration and runtime environments~\cite{haner2018software}. Thought with these differences, a common theme across definitions is the iterative interaction between classical and quantum components under hardware constraints, highlighting the need for system-level abstractions and engineering methodologies. 

Despite ongoing advances in quantum hardware and algorithms, the systematic engineering of hybrid quantum–classical applications remains underdeveloped. Current practices largely rely on low-level software development kits (SDKs) and ad hoc experimentation, making it difficult to assess whether a proposed solution is feasible under realistic hardware constraints such as limited qubit counts, restricted connectivity, noise, and execution overhead. This limitation is rooted in the intrinsic complexity of hybrid quantum–classical systems, which integrate heterogeneous computational components operating under probabilistic semantics and strict physical constraints. As a result, many candidate applications remain speculative and costly to evaluate. Existing efforts predominantly focus on algorithm design and programming abstractions, with limited support for system-level reasoning, design-space exploration, and early-stage feasibility assessment~\cite{leymannBitterTruthQuantum2020, wederQuantumSoftwareDevelopment2021, pezze20252030}.

To address this gap, this paper presents \qready, a vision for an MBSE-oriented framework that supports the systematic design and predictive feasibility assessment of hybrid quantum–classical applications. The core idea is to move beyond trial-and-error development toward a structured engineering process that enables early evaluation of candidate solutions. \qready adopts an MBSE perspective and uses MDE principles~\cite{incoseINCOSESystemsEngineering2023}, together with Quantum Software Engineering (QSE)~\cite{pezze20252030}, to organize the development process into a pipeline that spans requirements modeling, guided problem structuring, strategy configuration, executable workflow modeling, and feasibility assessment. By leveraging simulation and resource estimation, the framework aims to support the analysis and comparison of alternative designs prior to deployment on quantum hardware.

Rather than presenting a fully realized solution, \qready is intended as a research vision that highlights key challenges and outlines a potential direction for advancing the engineering foundations of quantum applications. To make the vision concrete, the paper develops a bank credit-portfolio capital-assessment example to illustarte the pipeline, from quantum-aware requirements to problem structuring, strategy instantiation, workflow modeling, and predictive feasibility decision-making. The example is not intended as a new financial-risk model or quantum algorithm, but to demonstrate how model-based traceability, supported by model-driven automation, can support early, evidence-based engineering decisions for a governed hybrid application.

The remainder of this paper is organized as follows: Section~\ref{sec:background} presents the required background. In Section~\ref{sec:qready}, we present details of our vision \qready. We conclude the paper in Section~\ref{sec:conclusion}.
\section{Background}
\label{sec:background}

\subsection{Hybrid Quantum-Classical Applications}
In the past decade, quantum software infrastructures have matured significantly. Tool chains such as IBM Qiskit~\cite{anisQiskitOpensourceFramework2021}, Origin Quantum QPanda~\cite{douQPandaHighperformanceQuantum2022}, and Microsoft Q\#~\cite{svoreEnablingScalableQuantum2018} provide high-level abstractions for developing variational and hybrid algorithms, while optimization-oriented platforms (e.g., QBoson and D-Wave) support Quadratic Unconstrained Binary Optimization (QUBO) modeling and annealing-based execution~\cite{zhangQuantumOptimizationSoftware2025,gloverTutorialFormulatingUsing2018}. In addition, cloud-based environments increasingly enable the orchestration of hybrid quantum–classical workflows across heterogeneous resources~\cite{gillQuantumCloudComputing2022,wederQuantumSoftwareDevelopment2021}. These developments indicate that quantum computing is evolving into a practical computational resource embedded within broader software ecosystems. Nevertheless, progress in hardware capabilities and algorithmic techniques has outpaced the development of system-level engineering methodologies, particularly for managing the complexity of hybrid applications.

Hybrid quantum–classical applications are inherently heterogeneous systems~\cite{fuQuingoProgrammingFramework2021,phillipsonClassificationHybridQuantumclassical2023}. They consist of multiple interacting components, such as classical preprocessing, problem encoding, quantum execution, feedback control loops, and post-processing. At the same time, these components operate under stringent constraints such as limited qubit resources, restricted connectivity, noise and decoherence, cloud latency, and stochastic measurement semantics. This combination of architectural heterogeneity and hardware-induced uncertainty makes it difficult to reason about system behavior and resource requirements at design time.

Such challenges become particularly critical in application domains where hybrid quantum–classical approaches are expected to provide advantages, such as scientific simulation, cryptographic analysis, large-scale optimization, and hybrid AI workflows~\cite{murilloQuantumSoftwareEngineering2025,leymannBitterTruthQuantum2020,destefanoSoftwareEngineeringQuantum2023,lewisFormalVerificationQuantum2023}. In these settings, practitioners must determine whether a given hybrid solution is feasible, scalable, and resource-efficient under realistic hardware constraints. However, due to the lack of systematic engineering support, such feasibility assessment remains largely ad hoc and difficult to perform prior to implementation. This motivates the need of \qready.

\subsection{Quantum Software Engineering (QSE)}
QSE adapts established software engineering principles to the development of quantum and hybrid quantum-classical systems~\cite{aliWhenSoftwareEngineering2022, pezze20252030}. While it draws on decades of classical software engineering experience, it must also account for distinctive quantum characteristics such as superposition and entanglement, as well as the practical limitations of NISQ hardware~\cite{preskillQuantumComputingNISQ2018}. Together, these features make system behavior harder to reason about at design time and strengthen the need for systematic engineering approaches that support early-stage feasibility reasoning and design-space exploration under realistic conditions.

As defined in ~\cite{pezze20252030}, QSE is ``an interdisciplinary field that focuses on the principles, methodologies, standards, and tools for designing, developing, testing, maintaining, and managing quantum software systems, including hybrid quantum-classic systems that run on quantum computers or quantum simulators'', and QSE ``involves the application of quantum-based techniques, including quantum algorithms such as quantum annealing, to solve complex problems in both classic and quantum software engineering''. 
Over the past few years, QSE has established its community and started to grow, marked by international conferences and workshops, dedicated journal special issues, and collaborative research initiatives. The QSE community has proposed various approaches tailored to deal with unique challenges in engineering quantum algorithms and applications. These methodologies span a broad spectrum, covering requirements engineering~\cite{spoletini2023towards, yueQuantumSoftwareRequirements2023, saraiva2021non}, quantum software architecture~\cite{aliModelingQuantumPrograms2020, gemeinhardtModelDrivenQuantumSoftware2021, wederQuantumSoftwareDevelopment2021, wederHybridQuantumApplications2021}, validating and verifying quantum software~\cite{li2026dynamic, mendiluzeMuskitMutationAnalysis2021, muqeetMitigatingNoiseQuantum2024, oldfieldFasterBetterQuantum2025, lewis2023formal}, etc. As the field continues to mature, the foundational work undertaken by the QSE community will undoubtedly play a pivotal role in shaping the future of quantum computing and real-world quantum applications. 

\subsection{Model-Based Systems Engineering, Model-Driven Engineering, and SysML}
Model-Based Systems Engineering (MBSE) has been widely adopted in classical domains to manage system complexity, support requirement traceability, and enable early validation~\cite{incoseINCOSESystemsEngineering2023}. Its emphasis on abstraction, formalization, and traceability makes it a promising foundation for addressing the engineering challenges of hybrid quantum–classical systems. Model-Driven Engineering (MDE) complements this perspective by providing transformation, automation, and analysis techniques over structured model artifacts. However, existing MBSE and MDE practices have largely been developed for classical systems and do not directly capture key characteristics of quantum applications, such as probabilistic execution semantics, hardware heterogeneity across quantum backends, topology-constrained qubit interactions, and encoding transformations from domain problems to quantum representations.

Consequently, a significant methodological gap persists: there is currently no unified MBSE approach, supported by MDE techniques, that integrates requirements engineering, hybrid architecture modeling, and hardware-aware feasibility analysis within a coherent systems engineering framework. This paper addresses this gap by proposing \qready, an MBSE-oriented approach that applies MDE principles to hybrid quantum–classical systems and enables executable, backend-aware system modeling for predictive feasibility assessment.

The Systems Modeling Language (SysML) has long served as a standard modeling language in MBSE, widely adopted across domains such as aerospace, automotive, and cyber-physical systems. Standardized by the Object Management Group (OMG), it provides a general-purpose notation for representing system requirements, structure, and behavior within a unified framework. SysML v2 is the new generation of this systems modeling language for MBSE and a strong foundation for MDE-style automation over system models. It introduces a more rigorous semantic foundation, grounded in the Kernel Modeling Language (KerML), and improves modularity, reuse, variability, and constraint specification~\cite{OMG-KerML-1.0,OMG-SysML-2.0}. Importantly, SysML v2 is no longer only a diagram-oriented notation: its concrete syntax includes both textual and graphical forms, enabling the same model to be manipulated as structured text and inspected through stakeholder-oriented views. Industrial tool support is also emerging; for example, Dassault Systèmes provides CATIA SysML v2 within the CATIA Magic/No Magic tooling ecosystem, with support for synchronization between textual and graphical representations~\cite{dassaultCATIASysMLv2}.

These characteristics make SysML v2 particularly suitable for \qready. The textual notation provides machine-readable model artifacts that can be parsed, transformed, versioned, checked, and used as context for large language model (LLM)-based modeling assistance. The graphical notation complements this by supporting human comprehension, communication, and review of requirements, architectures, workflows, and feasibility evidence. This dual representation is important because \qready aims to support both automated reasoning and human-in-the-loop engineering decisions. More broadly, the potential of integrating LLMs into MDE has already been recognized by the MDE community~\cite{LLM4MDESurvey}, further motivating an MBSE foundation in SysML v2 that exposes models in forms usable by both engineers and AI-assisted tooling.

\section{\qready}
\label{sec:qready}
In this section, we present our vision \qready in detail. Section~\ref{subsec:overview} gives an overview of the framework, and Section~\ref{subsec:runningExample} introduces the running credit-portfolio example used throughout the paper. Sections~\ref{subsec:QRE}--\ref{subsec:feasibilityAssessment} then present the main \qready components one by one. For each component, we discuss the research rationale, summarize the relevant state of the art, outline the research roadmap, and illustrate the component through the running example. Finally, Section~\ref{sec:exampleSynthesis} synthesizes the running example across the complete \qready pipeline.

\subsection{Overview}
\label{subsec:overview}
\begin{figure}[htbp]
	\centering
	\includegraphics[width=\linewidth]{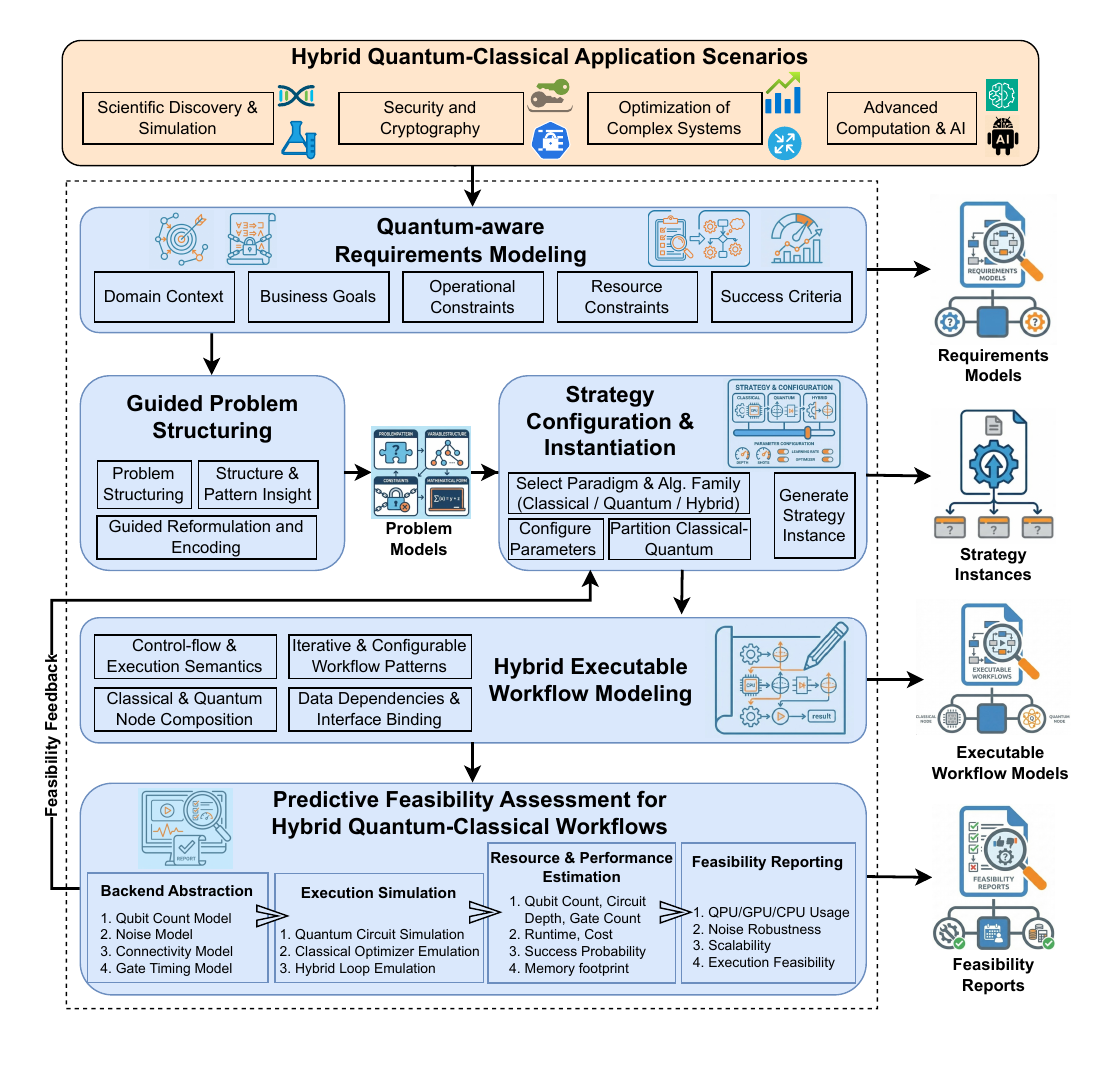}
		\caption{Overview of the \qready framework. The pipeline starts with quantum-aware requirements modeling, transforms requirements into structured computational problem models, derives validated hybrid strategy instances, converts them into executable workflow models, and evaluates candidate workflows through predictive feasibility assessment. Feedback links support iterative refinement when constraints, backend assumptions, or feasibility results change.}
		\Description{Overview diagram of \qready showing the main pipeline components: quantum-aware requirements modeling, guided problem structuring, strategy configuration and instantiation, hybrid executable workflow modeling, and predictive feasibility assessment, with feedback links for iterative refinement.}
	\label{fig:overview}
\end{figure}

As shown in Figure~\ref{fig:overview}, the \qready framework is organized into five tightly connected components that together form a coherent engineering pipeline. The pipeline begins with \textit{Quantum-aware Requirements Modeling (QRE)}, which establishes a structured requirements foundation for later analysis. It then proceeds to \textit{Guided Problem Structuring}, which transforms formalized requirements into computational problem models suitable for hybrid quantum-classical solution design, and to \textit{Strategy Configuration \& Instantiation}, which defines mechanisms for generating alternative hybrid solution strategies from those abstractions through variation points, parameter spaces, and constraint rules. Next, \textit{Hybrid Executable Workflow Modeling} converts strategy instances into formally defined executable workflows that can be directly analyzed. Finally, \textit{Predictive Feasibility Assessment} evaluates candidate workflows using backend abstractions, simulation techniques, and analytical estimators, producing structured feasibility results that support comparison and iterative refinement.

\qready is envisioned to contribute in four main ways: (1) framing hybrid quantum-classical applications from an MBSE perspective; (2) providing a unified MBSE-based framework for configuration and predictive feasibility assessment, grounded in MDE principles; (3) enabling traceable system-level models that span strategies, workflows, backend constraints, and data dependencies; and (4) supporting hardware-aware predictive analysis through backend abstractions and noise-aware modeling. Together, these contributions define the methodological direction of the vision and aim to strengthen the foundations of QSE for hybrid quantum-classical applications.
Note that \qready does not rely on the availability of specific quantum devices. Instead, it provides structured modeling, configuration, execution emulation, and quantitative assessment capabilities that remain adaptable to evolving hardware landscapes. When quantum hardware becomes accessible, workflows can be deployed for empirical validation. The core value of the framework lies in enabling informed, evidence-based feasibility reasoning before deployment decisions are made.

Across the pipeline, LLM-assisted modeling can support elicitation, guidance, model completion, and explanation~\cite{LLM4MDESurvey}. However, \qready treats SysML v2 models, explicit constraints, and quantitative feasibility checks as the authoritative artifacts used for validation and decision-making. This division of roles allows LLMs to assist users without displacing formal model semantics or evidence-based feasibility reasoning.

\subsection{Running Example: Bank Credit-Portfolio Capital Assessment}
\label{subsec:runningExample}
We use a bank’s credit-portfolio risk system as a running example. The system supports a concrete business decision: determining how much financial reserve the bank should hold to remain solvent if an unusually large number of borrowers fail to repay their loans over the next year. Every quarter, risk analysts execute an approved software workflow that reads the bank’s loan portfolio, simulates many possible economic scenarios, estimates the resulting credit losses, and submits the results for independent model-risk review. Capital planners then use the approved outputs to set financial reserves and lending limits.

Each loan record provides three principal inputs. \emph{Exposure at default (EAD)} is the amount still owed to the bank if the borrower defaults. \emph{Probability of default (PD)} estimates the likelihood that the borrower will default within one year. \emph{Loss given default (LGD)} is the fraction of the EAD that the bank does not expect to recover after taking into account collateral, repayments, and other recoveries. For example, if a \$1 million loan has an LGD of 40\%, it produces a \$400,000 loss in the event of default. In the remainder of this example, we use the term \emph{facility} to refer to an individual loan or credit agreement, and \emph{segment} to refer to a group of facilities with similar risk characteristics.

\begin{table}[htbp]
	\centering
	\footnotesize
	\caption{Synthetic credit portfolio used in the running example}
	\label{tab:creditInput}
	\begin{tabular}{lrrrr}
		\toprule
		Portfolio segment & Loans & Amount owed (\$m) & Default chance & Loss share \\
		\midrule
		Large corporate & 600 & 300 & 0.8\% & 45\% \\
		Mid-market & 1,400 & 250 & 1.5\% & 40\% \\
		SME & 3,000 & 180 & 2.5\% & 35\% \\
		Commercial real estate & 500 & 120 & 1.2\% & 30\% \\
		Retail secured & 2,500 & 100 & 1.0\% & 20\% \\
		Retail unsecured & 2,000 & 50 & 4.0\% & 65\% \\
		\midrule
		Total & 10,000 & 1,000 & & \\
		\bottomrule
	\end{tabular}
\end{table}

The concrete context is a quarterly assessment of the synthetic 10,000-loan, \$1 billion portfolio summarized in Table~\ref{tab:creditInput}. Defaults cannot be modeled as unrelated events because an economic downturn can cause many borrowers to default together. The approved classical system therefore generates correlated default scenarios and calculates the total loss in each scenario. From the resulting loss distribution, it reports: (1) \emph{expected loss}, the average loss across scenarios; (2) \emph{99.9\% Value-at-Risk (VaR)}, a loss threshold exceeded in approximately one out of every 1,000 modeled years; and (3) \emph{economic capital}, the additional reserve between expected loss and that severe-loss threshold.

The computationally expensive step is repeatedly estimating rare-event probabilities from a large number of simulated scenarios. Quantum amplitude estimation (QAE) is a family of quantum algorithms proposed for estimating such probabilities using fewer model evaluations than classical Monte Carlo under suitable assumptions~\cite{woernerQuantumRiskAnalysis2019,eggerCreditRiskAnalysis2021}. In software-engineering terms, the candidate quantum component would replace one performance-critical probability-estimation service inside an existing governed workflow; it would not replace the complete risk system. The engineering question for \qready is whether this substitution can satisfy accuracy, runtime, security, auditability, and hardware constraints. Although the portfolio data are synthetic, the actors, workflow, inputs, outputs, and governance concerns reflect a real banking application. The example is not intended to contribute a new financial-risk model or quantum algorithm; it demonstrates the software-engineering challenges of introducing a probabilistic, hardware-constrained quantum service into an existing governed software system.

\subsection{Quantum Requirements Modeling}
\label{subsec:QRE}

\subsubsection{Research Rationale}
MBSE languages and notations such as SysML provide established mechanisms for specifying system boundaries, stakeholder goals, constraints, and verification relations. However, a hybrid quantum--classical application introduces an additional requirements-engineering question: not only \emph{what} the system must achieve, but also \emph{which obligations belong to the classical part, the quantum part, or their interaction}. This allocation matters because system-level qualities such as accuracy and runtime may depend on quantum-specific quantities including logical and physical qubit demand, circuit depth, gate count, sampling budget, noise, and backend availability.

These concerns must be represented before algorithm and backend selection. Otherwise, an attractive system-level goal can remain disconnected from the resource assumptions that determine whether it is realizable. Prior QSE research similarly emphasizes portability across heterogeneous quantum platforms, robustness under noise, scalability under constrained qubit resources, and explicit quantum--classical separation~\cite{sepulvedaSystematicReviewRequirements2024,hacalogluExploratoryReviewQuantum2024}. We therefore treat quantum requirements modeling as the creation of a traceable argument from stakeholder intent to measurable hybrid, classical, and quantum obligations that can guide design-space exploration and later feasibility assessment.

\subsubsection{State of the Art}
Saraiva et al.~\cite{saraiva2021non} showed that quality and constraint concerns such as performance, reliability, scalability, and maintainability are central, hardware-dependent concerns for quantum programs rather than secondary considerations. Spoletini~\cite{spoletini2023towards} examined how established requirements-engineering activities, including elicitation, analysis, specification, and validation, need to be adapted for quantum software.

Pérez-Castillo and Piattini~\cite{perez2022design} proposed unified modeling language (UML) stereotypes to explicitly capture quantum-related concerns in use case diagrams. Specifically, they introduced the «Quantum» stereotype to identify use cases involving quantum computation (e.g., quantum chemistry), defined a «Quantum Computer» actor to represent quantum execution resources, and used the «Quantum Request» stereotype on include relationships to denote invocations of quantum software components.
Yue et al.~\cite{yueQuantumSoftwareRequirements2023} proposed a SysML requirements diagram for credit-risk analysis that uses «cReq», «qReq», and «hReq» to identify classical, quantum, and hybrid responsibility. Importantly, the diagram also relates high-level performance expectations to estimable quantities such as required gates and ancilla qubits, and then to hardware constraints such as available qubits and circuit depth. This illustrates that quantum-aware requirements modeling is not merely the addition of quantum labels: it must expose decomposition, dependency, and allocation relations that make feasibility assumptions analyzable.

Existing proposals establish useful notations, but they remain preliminary and provide limited methodological and tool support for maintaining these relations through later engineering phases. In particular, a requirements model must support systematic refinement from stakeholder goals to measurable acceptance criteria, allocation to hybrid components, explicit assumptions about candidate backends, and traceability to the evidence used for verification.

\subsubsection{Research Roadmap}
First, \qready introduces hybrid-specific extensions to SysML v2 requirements notations to support the structured specification of hybrid quantum-classical applications. Building on~\cite{yueQuantumSoftwareRequirements2023}, each requirement can be marked as \texttt{cReq}, \texttt{qReq}, or \texttt{hReq} to indicate whether responsibility lies with the classical part, the quantum part, or the hybrid system. A high-level hybrid requirement can therefore be decomposed into classical and quantum subrequirements without losing its original stakeholder rationale. The library also defines typed relations for decomposition, satisfaction, constraint, dependency, and verification.

Second, \qready supports the structured elicitation of domain context, stakeholders, system boundaries, goals, operational assumptions, and the potential value of quantum integration. Guided templates of \qready should clarify the task to be achieved, why a quantum component is being considered, which classical baseline remains authoritative, and under which conditions the hybrid system operates. At this stage, the objective is to establish the problem and its acceptance conditions, not to commit prematurely to a particular mathematical formulation, quantum algorithm, or backend.

Third, \qready refines stakeholder goals into measurable requirements and exposes their feasibility dependencies. System-level success criteria, such as an accuracy or runtime target, are related to quantum resource obligations such as estimated qubits, gates, circuit depth, shots, and error budget, as well as to backend constraints and classical orchestration costs. These values may initially be unknown; the requirements model records them as quantities to be estimated and binds them to explicit verification methods and evidence. This allows later phases to populate the estimates and determine whether a candidate strategy satisfies the original requirements.

Finally, the requirements model serves as a live traceability backbone for the remaining \qready pipeline. Problem formulations trace to the requirements they refine; strategy configurations are rejected when they violate allocation or resource constraints; executable workflows identify the components that satisfy each requirement; and feasibility results provide verification evidence. This continuity turns the requirements diagram from descriptive documentation into an analyzable contract for design and assessment.

\subsubsection{Running Example}
\begin{figure*}[htbp]
	\centering
		\includegraphics[width=\linewidth]{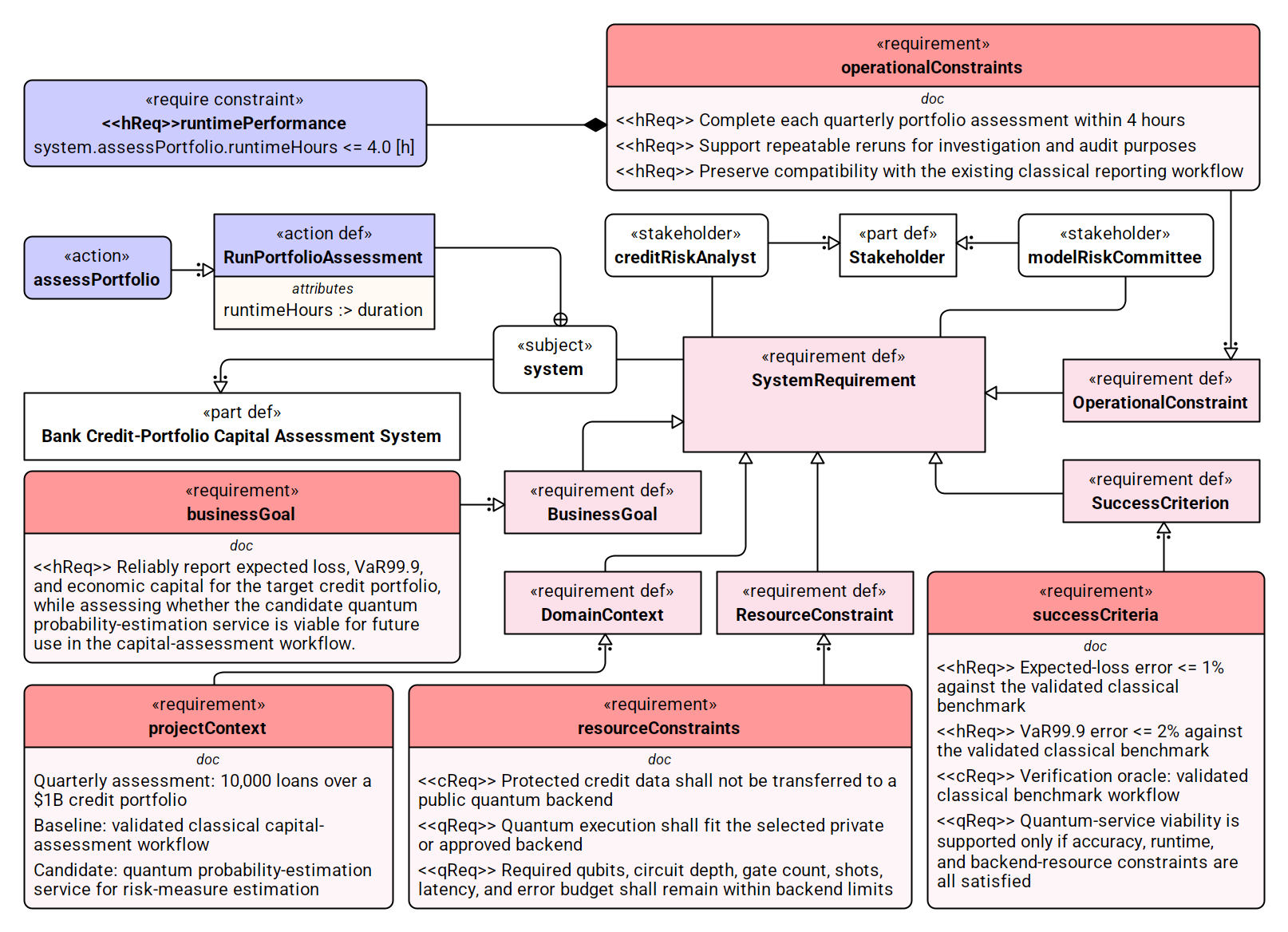}
			\caption{SysML v2 requirements diagram for the credit-risk example. The pink elements define five specialized types of \texttt{SystemRequirement}: \texttt{DomainContext}, \texttt{BusinessGoal}, \texttt{OperationalConstraint}, \texttt{ResourceConstraint}, and \texttt{SuccessCriterion}. The diagram also defines the subject of \texttt{SystemRequirement}, namely the \texttt{Bank Credit-Portfolio Capital Assessment System}, including its \texttt{assessPortfolio} action and the \texttt{runtimeHours} attribute used for runtime reasoning. The requirement definition is associated with two stakeholder types, \texttt{creditRiskAnalyst} and \texttt{modelRiskCommittee}. The blue elements illustrate that a textual requirement can be formalized as an executable constraint: the first \texttt{<<hReq>>} in \texttt{operationalConstraints} is represented by the \texttt{runtimePerformance} constraint over \texttt{system.assessPortfolio.runtimeHours}.}
			\Description{A SysML v2 requirements diagram for the credit-risk example. Pink elements show five specialized SystemRequirement types: DomainContext, BusinessGoal, OperationalConstraint, ResourceConstraint, and SuccessCriterion. The diagram defines the subject as the Bank Credit-Portfolio Capital Assessment System with an assessPortfolio action and runtimeHours attribute. It associates SystemRequirement with creditRiskAnalyst and modelRiskCommittee stakeholders. Blue elements show that the first hReq in operationalConstraints is formalized as a runtimePerformance constraint over system.assessPortfolio.runtimeHours.}
	\label{fig:creditRequirements}
\end{figure*}

The QRE component of \qready organizes the running example as the SysML v2 requirements model shown in Figure~\ref{fig:creditRequirements}. The model introduces a reusable \texttt{SystemRequirement} requirement definition whose subject is a \texttt{Bank Credit-Portfolio Capital Assessment System}. Within this subject, the model defines a \texttt{PortfolioAssessmentRun} action with a \texttt{runtimeHours} duration attribute, and instantiates the system action \texttt{assessPortfolio}. The same reusable requirement definition also declares the two key stakeholders, \texttt{creditRiskAnalyst} and \texttt{modelRiskCommittee}. The five \qready requirement-model elements are then defined as specializations of \texttt{SystemRequirement}: \texttt{DomainContext}, \texttt{BusinessGoal}, \texttt{OperationalConstraint}, \texttt{ResourceConstraint}, and \texttt{SuccessCriterion}.

The \texttt{projectContext} instance of \texttt{DomainContext} records the quarterly assessment of a synthetic 10,000-loan, \$1 billion portfolio, the validated classical capital-assessment workflow, and the candidate quantum probability-estimation service for risk-measure estimation. From this context, the \texttt{businessGoal} instance states the stakeholder value to be delivered: reliably report expected loss, 99.9\% VaR, and economic capital while assessing whether the candidate quantum service is viable for future use in the capital-assessment workflow.

The \texttt{operationalConstraints} requirement contains hybrid operational obligations, including completion of each quarterly assessment within 4 hours, repeatable reruns for audit and investigation, and compatibility with the existing classical reporting workflow. Its \texttt{runtimePerformance} constraint formalizes the runtime requirement as \texttt{system.assessPortfolio.runtimeHours <= 4.0 [SI::h]}.
The \texttt{resourceConstraints} requirement captures both classical and quantum resource limits: protected credit data must not be transferred to a public quantum backend, quantum execution must fit the selected private or approved backend, and required qubits, circuit depth, gate count, shots, latency, and error budget must remain within backend limits. The \texttt{successCriteria} requirement specifies measurable acceptance conditions: expected-loss error below 1\% and VaR99.9 error below 2\% against the validated classical benchmark, with quantum-service viability supported only when accuracy, runtime, and backend-resource constraints are all satisfied.

Within these five \qready model elements, we adopt the notation introduced in~\cite{yueQuantumSoftwareRequirements2023} to distinguish responsibility across the hybrid system. The stereotypes «cReq», «qReq», and «hReq» indicate requirements for the classical part, quantum part, and hybrid system, respectively. Thus, the business goal, runtime obligation, compatibility obligation, and accuracy thresholds are \texttt{hReq}s; the validation oracle and protected-data restriction include \texttt{cReq} responsibilities; and backend-capacity and quantum-service viability obligations are \texttt{qReq}s.

These elements make feasibility reasoning explicit already at the requirements level. Figure~\ref{fig:creditRequirements}, for example, does not leave the runtime obligation as informal prose; it formalizes it as the \texttt{runtimePerformance} constraint over \texttt{system.assessPortfolio.runtimeHours}. More generally, the same requirements model keeps accuracy thresholds, validation baselines, protected-data restrictions, and backend-capacity limits explicit, so that later strategy configuration can reject inadmissible candidates before implementation. In this way, the requirements model serves not only as documentation of stakeholder intent but also as a typed basis for downstream feasibility assessment.

\subsection{Guided Problem Structuring}
\label{subsec:ProblemStructuring}
\subsubsection{Research Rationale}
Computational structuring forms the critical bridge between formalized requirements (Section~\ref{subsec:QRE}) and systematic strategy configuration (Section~\ref{subsec:strategyConfigInstantiation}).
Optimization theory provides mature foundations for transforming domain problems into well-established solver-oriented formulations, such as linear programming models and quadratic binary formulations, including Quadratic Unconstrained Binary Optimization (QUBO)~\cite{gloverTutorialFormulatingUsing2018} and Ising models~\cite{lucasIsingFormulationsMany2014}. Computational complexity theory further supports the characterization of these formulations in terms of problem classes, reductions, and expected computational difficulty. Nevertheless, the translation from domain-level descriptions to problem models suitable for hybrid quantum-classical workflows remains largely expert-driven~\cite{babbushGrandChallengeQuantum2025}.

Beyond optimization, recent discussions in QSE also highlight the absence of systematic methods for early-stage problem structuring, structural pattern recognition, and guided reformulation that preserve domain semantics while preparing problems for quantum suitability analysis~\cite{murilloQuantumSoftwareEngineering2025, wederQuantumSoftwareDevelopment2021}. In practice, domain experts must manually interpret whether a problem is combinatorial, sampling-based, linear-algebraic, or variational in nature, and whether its structure aligns with qubit-based representations or hybrid decomposition strategies. This reliance on tacit expertise creates a significant barrier to adoption and limits reproducibility~\cite{babbushGrandChallengeQuantum2025,murilloQuantumSoftwareEngineering2025,wederQuantumSoftwareDevelopment2021}. Although algorithmic paradigms such as variational quantum algorithms and hybrid optimization loops are now well established~\cite{VariationalQuantumAlgorithms2021,bhartiNoisyIntermediatescaleQuantum2022}, mainstream software platforms still focus primarily on circuit construction, execution, and algorithm libraries, with comparatively limited support for structured, traceable reformulation from high-level problem descriptions~\cite{gemeinhardtModelDrivenQuantumSoftware2021,murilloQuantumSoftwareEngineering2025,wederQuantumSoftwareDevelopment2021}.

\subsubsection{State of the Art}
Zhang et al. in their recent survey~\cite{zhangQuantumOptimizationSoftware2025, zhangEmpiricalStudiesQuantum2025} analyzed quantum, quantum-inspired, and hybrid optimization, identifying recurring structural types (primarily QUBO and Ising models) alongside gaps in formulation rigor and standardized practices. While emerging tools like LLM-QUBO~\cite{pauckertLLMQUBOEndtoEndFramework2025} attempt to automate the mathematical transformation from natural language to QUBO matrices, bridging the gap from high-level application (domain) requirements to structured problem models remains largely infeasible without manual expert intervention. Furthermore, QUBO/Ising represent only one paradigm; other critical models, such as gate-based circuits and variational quantum circuits (VQCs), require entirely different formalization and execution semantics. 
These findings directly motivate guided problem structuring, which replaces today’s ad hoc, expert-driven translations with a systematic, education-oriented MBSE environment that leverages a curated computational pattern library and MDE-style guided reformulation to help users transform requirement-level descriptions into explicit, hybrid-aware problem models. In this setting, LLM assistance can help interpret domain descriptions, suggest candidate structural patterns, and explain reformulation alternatives, while the resulting problem model remains a typed and reviewable artifact rather than an unconstrained natural-language output.

\subsubsection{Research Roadmap}
First, guided templates help clarify decision objectives (e.g., optimization, estimation, simulation), variable types (binary, integer, continuous), constraint categories, system scale, and expected outputs, instead of requiring formal mathematical derivations. Importantly, the problem structuring process captures problem-domain characteristics that influence potential quantum realizability, without yet committing to specific algorithms or circuits. For example, it identifies whether the problem is predominantly discrete (suggesting natural qubit-based representations), whether it exhibits strong combinatorial coupling between variables (which may translate into multi-qubit interactions), and how the underlying dependency structure can be represented as a graph (e.g., sparse, clustered, or densely connected). The intent is to remain at the level of problem semantics while encoding information that guides encoding choices and hybrid decomposition. 

Second, a compact and well-structured computational pattern library captures recurring structural patterns relevant to hybrid quantum-classical applications. These patterns are systematically collected and distilled from the scientific literature, including established quantum algorithm classes, hybrid variational schemes, and commonly studied optimization and simulation formulations. The library organizes stable problem forms such as combinatorial optimization, quadratic binary models (e.g., QUBO/Ising-type structures), sampling tasks, linear-algebraic kernels, and variational optimization loops. Each pattern is characterized by its defining structural properties, interaction topology, scaling tendencies, and typical computational implications. Based on this curated body of knowledge, the system generates structure pattern insights for end users by matching their structured problem models to relevant archetypes. The feedback provides interpretable explanations of structural similarities, expected scaling behavior, coupling density, and indicative computational hardness.

Third, guided reformulation mechanisms assist users in progressively mapping structured problem models toward computationally explicit representations. Building on the outputs of the previous tasks, modeling templates and stepwise guidance support the exploration of alternative reformulation pathways, including classical formulations and hybrid-compatible encodings. The process is explicitly human-in-the-loop and explanatory: proposed transformations are presented as structured alternatives, accompanied by rationale and implications, rather than enforced or automatically optimized conversions. LLM-generated explanations can make these alternatives easier to inspect, but admissibility and traceability are determined by the structured model and its constraints. To support these choices, the framework also provides approximate indicators of representation size, structural growth, variable expansion, and resource implications, such as potential qubit scaling or interaction amplification. These indicators help users compare reformulation alternatives in a transparent and traceable way, while keeping the focus on guided transformation rather than automated encoding generation. 

\subsubsection{Running Example}
Guided problem structuring starts from the QRE model in Section~\ref{subsec:QRE}, rather than from an informal problem statement, and produces the structured problem model consumed by Strategy Configuration and Instantiation (Section~\ref{subsec:strategyConfigInstantiation}). In the credit-risk example, the QRE elements provide the input boundary: \texttt{projectContext} identifies the target portfolio, validated classical benchmark, and candidate quantum probability-estimation service; \texttt{businessGoal} identifies the risk measures to preserve; and \texttt{operationalConstraints}, \texttt{resourceConstraints}, and \texttt{successCriteria} define the runtime, backend, data-protection, and accuracy obligations that later strategy choices must satisfy.

\textit{Problem Structuring.} \qready first extracts the computational task, inputs, outputs, and acceptance obligations from the QRE model. It creates \texttt{CreditRiskProblem\_01} and classifies it as a \emph{probabilistic risk-estimation task}, because expected loss, VaR99.9, and economic capital are derived from the portfolio loss distribution rather than from an optimization objective. The structured data flow is recorded as approved facility records $\rightarrow$ conditional default probabilities $\rightarrow$ loss distribution $\rightarrow$ cumulative distribution $F_L(x)=P[L\leq x]$ $\rightarrow$ bisection search for the smallest $x$ with $F_L(x)\geq0.999$. This step preserves traceability to the QRE \texttt{businessGoal} and \texttt{successCriteria}: the problem model must produce expected loss and VaR99.9, and the resulting approximation error must later be checked against the 1\% and 2\% thresholds.

\textit{Structure and Pattern Insight.} \qready then characterizes the structural properties that influence downstream strategy selection. The model records that borrower defaults are correlated, that the central computational object is the portfolio loss distribution together with a rare-tail probability query for VaR estimation, that the input scale is 10,000 loans over a \$1 billion portfolio, and that direct facility-level encoding may create a large binary state space. These descriptors characterize the problem as a probabilistic portfolio-loss estimation task rather than an optimization task. They also indicate that classical Monte Carlo simulation is the natural benchmark for estimating the loss distribution, while the rare-tail probability query provides a plausible entry point for a quantum amplitude-estimation strategy, subject to state-preparation and encoding feasibility. Concretely, the problem requires simulating uncertain and correlated borrower defaults, constructing or approximating the resulting portfolio loss distribution, and repeatedly estimating probabilities for candidate VaR thresholds, either in cumulative form $F_L(x)=P[L\leq x]$ or equivalently in tail form $P[L>x]$. In this sense, the natural computational structure is probability estimation over a loss distribution, not binary objective-function minimization. This explains why a premature QUBO reformulation would obscure the actual credit-risk analysis task.

\textit{Guided Reformulation and Encoding.} Finally, \qready explores a small set of candidate reformulations for the same structured problem model rather than committing directly to a quantum algorithm. Here, \texttt{CreditRiskProblem\_01} remains the stable problem-level artifact, while \texttt{F1-FacilityLevel}, \texttt{F2-SegmentAggregated}, and \texttt{F3-HardwarePilot} represent alternative formulation paths attached to that problem model. These three candidates are not arbitrary: they are produced by guided transformations over the same portfolio-loss model, namely preserving the original facility-level granularity, aggregating facilities into the six business segments already present in the input model, and restricting the model to a small representative subset for hardware-oriented feasibility exploration. Rather than expressing the outcome as a detailed formulation listing, \qready can present the decision as a compact evaluation template that compares candidate formulations along a small set of selection criteria. In this template, \emph{fidelity} denotes how closely a formulation preserves the original domain semantics, \emph{abstraction} captures the degree of aggregation introduced during modeling, \emph{traceability} indicates how directly modeling decisions can be linked back to the QRE, and \emph{resource fit} summarizes expected compatibility with the stated execution constraints. This should be read only as a running example: in the full \qready approach, richer templates with additional aspects are expected to support a broader range of problem-modeling situations. Table~\ref{tab:reformulationPaths} shows this template for the running example.

\begin{table}[t]
\centering
\footnotesize
\caption{Comparison of reformulation alternatives for the running example}
\label{tab:reformulationPaths}
\setlength{\tabcolsep}{3pt}
\renewcommand{\arraystretch}{1.06}
\begin{tabular}{@{}lccccc@{}}
\toprule
\textbf{Formulation} & \textbf{Fidelity} & \textbf{Abstraction} & \textbf{Traceability} & \textbf{Resource fit} & \textbf{Selected} \\
\midrule
\shortstack[l]{\texttt{F1-FacilityLevel}} & High & Low & High & Low & No \\
\shortstack[l]{\texttt{F2-SegmentAggregated}} & Medium & Medium & Medium & High & Yes \\
\shortstack[l]{\texttt{F3-HardwarePilot}} & Low & High & Low & High & No \\
\bottomrule
\end{tabular}
\end{table}

For the running example, the template marks \texttt{F2-SegmentAggregated} as the selected option because it provides the best balance between domain fidelity, manageable abstraction, retained traceability, and resource fit under the stated constraints. In this presentation style, the handoff to Section~\ref{subsec:strategyConfigInstantiation} is captured as a structured selection summary: one formulation is retained for the next phase, while the remaining candidates stay documented as rejected but traceable alternatives.


\subsection{Strategy Configuration and Instantiation}
\label{subsec:strategyConfigInstantiation}
\subsubsection{Research Rationale}
Current practice in selecting classical, quantum, or hybrid computational strategies remains largely heuristic and expert-driven. Paradigm selection, algorithm choice, hybrid partitioning, and parameter tuning are typically performed manually, often guided by informal reasoning or tool-specific defaults, with limited formal linkage to the structural characteristics of the underlying problem model~\cite{leymannBitterTruthQuantum2020, bhartiNoisyIntermediatescaleQuantum2022, VariationalQuantumAlgorithms2021}. Even when configuration tools are available, they frequently operate as menu-based selectors without explicit semantic constraints, compatibility validation, or traceable justification~\cite{laroseOverviewComparisonGate2019}. This lack of structured configuration logic reduces reproducibility, complicates systematic exploration of alternatives, and weakens continuity toward executable workflow generation (Section~\ref{subsec:exeWorkflow}) and feasibility analysis (Section~\ref{subsec:feasibilityAssessment}).

\subsubsection{State of the Art}
Extensive experience in classical domains demonstrates that formal specification techniques, such as the Object Constraint Language (OCL), are effective in capturing complex domain constraints in an executable and analyzable form, while automation (e.g., with search-based methods) enables systematic exploration of configuration spaces~\cite{lu2017automated, sartaj2025search, steimann2025meet}. These approaches rely fundamentally on appropriate abstractions and modeling support, suggesting that established principles from classical software engineering can inform systematic modeling, configuration, and validation in hybrid quantum–classical systems.

In this context, classical software engineering has accumulated decades of experience in modeling and metamodeling, resulting in a rich ecosystem of languages and tools. Widely adopted notations such as UML, SysML, and Business Process Model and Notation (BPMN), together with platforms such as Papyrus, Eclipse Modeling Framework (EMF), and PlantUML, provide mature foundations for abstraction, analysis, and model-driven development. Building on these foundations, recent work has begun to extend modeling techniques to the quantum domain. For example, Pérez-Castillo and Piattini~\cite{perez2022design} introduce UML stereotypes to capture quantum-related concerns at the requirements level, while Guo et al.~\cite{guo2025quanuml} propose QuanUML, which incorporates constructs such as qubits and quantum gates into UML-based models.

Beyond UML-based extensions, a range of representations has been explored for quantum software, including cause–effect graphs for testing~\cite{oldfieldInvestigatingQuantumeffect2022}, quantum flowcharts for algorithm-level control and data flow~\cite{QPL}, and quantum decision diagrams for representing and analyzing quantum states and circuits~\cite{wille2022decision}. At a lower level of abstraction, quantum circuit diagrams remain the dominant representation, describing sequences of quantum gates applied to qubits and supporting various optimizations such as circuit depth reduction and noise-aware transformations~\cite{9384317,10.1145/3565271,10.1145/3539613}. 

However, these representations primarily operate at the level of notation or algorithmic structure and provide limited support for system-level abstraction. Although tools such as Qiskit, Microsoft's Quantum Development Kit (QDK), and PennyLane enable modeling and simulation of quantum circuits, and frameworks such as the MQT DD Package~\cite{zulehner2019package} support decision-diagram-based representations, they largely focus on execution and analysis rather than enabling higher-level reasoning about alternative system designs. As a result, the abstraction level remains insufficient for supporting early-stage modeling, design-space exploration, and feasibility analysis of hybrid quantum–classical applications.

Therefore, modeling notations and tools alone are insufficient to support effective abstraction. Each abstraction must serve a well-defined analytical or engineering purpose. In particular, there is currently a lack of methodologies that systematically scope and constrain the strategy configuration space of hybrid quantum–classical applications, limiting the ability to reason about feasible design alternatives in a principled manner.

\subsubsection{Research Roadmap}
\qready addresses this gap through an MBSE-oriented, hence structured configuration space for hybrid computational strategies, grounded in SysML v2 and supported by MDE-style transformations. Strategy selection is treated not as ad hoc option picking, but as a constrained configuration process defined over a structured metamodel. SysML v2 provides typed relationships, variability mechanisms, and integrated constraint support, enabling strategy elements such as computational paradigm, algorithm family, partitioning structure, and parameters, to be explicitly modeled and semantically connected to the problem structures (Section~\ref{subsec:ProblemStructuring}). Metamodel constraints encode compatibility and dependency rules derived from algorithmic assumptions, hardware constraints, and emerging best practices. Overall, the enriched problem models produced in the \textit{Guided Problem Structuring} phase are transformed into validated strategy instances represented as SysML v2 artifacts. Each strategy instance captures a coherent and constraint-validated configuration of paradigm, algorithmic approach, hybrid structure, and parameters, with explicit traceability to the originating problem structure, thereby establishing a rigorous bridge between problem abstraction and executable workflow modeling (Section~\ref{subsec:exeWorkflow}).

The first task establishes a strategy configuration metamodel, implemented as an extension library within SysML v2, to ensure integration with the existing modeling environment. The full library may contain many reusable strategy options, while Listing~\ref{lst:strategyMetamodel} presents only the core excerpt needed to explain the running example. In this excerpt, a \texttt{StrategyInstance} links back to a structured \texttt{ProblemModel}, binds the main variation points considered here, namely paradigm, algorithm, encoding, and backend, and stores the feasibility quantities used for validation. The enumerations are included because they define the controlled value spaces referenced by the compatibility rules.
The omitted parts of the metamodel define concrete option elements, such as \texttt{HybridQuantumClassical} and \texttt{ClassicalOnly} for execution paradigms, \texttt{IterativeAmplitudeEstimation} and \texttt{QAOA} for algorithmic choices, and \texttt{CreditLossDistributionEncoding} and \texttt{QUBOProblemEncoding} for encoding choices. These elements are treated as members of an option library rather than as the conceptual core of the metamodel. The aim is to establish a controlled and analyzable variability backbone that captures the admissible configuration space while remaining natively embedded within SysML v2.

\begin{lstlisting}[language=,caption={Textual SysML v2 excerpt of the core strategy-configuration metamodel},label={lst:strategyMetamodel},basicstyle=\scriptsize\ttfamily,numbers=none,aboveskip=2pt,belowskip=2pt]
package QREADY_StrategyConfigurationMetamodel {
  part def ProblemModel {
    attribute problemId : String;
    attribute problemType : ProblemType;
    attribute structuringObjective : String;
  }

  part def StrategyInstance {
    attribute strategyId : String;
    attribute status : ValidationStatus;
    ref problem : ProblemModel;
    part paradigm : ComputationalParadigm;
    part algorithm : AlgorithmFamily;
    part encoding : ProblemEncoding;
    part backend : BackendProfile;
    part evaluation : FeasibilityEvaluation;

    constraint validProbabilityStrategy {
      problem.problemType == ProblemType::ProbabilisticRiskEstimation implies
      algorithm.supportsProbabilityEstimation == true
    }
    constraint problemModelEncodingIncompatibility {
      problem.problemType == ProblemType::ProbabilisticRiskEstimation implies
      encoding.encodingType != EncodingType::QUBOEncoding
    }
  }

  part def ComputationalParadigm {
    attribute name : String;
    attribute isQuantumEnabled : Boolean;
    attribute isHybrid : Boolean;
  }

  part def AlgorithmFamily {
    attribute name : String;
    attribute algorithmPattern : AlgorithmPattern;
    attribute supportsProbabilityEstimation : Boolean;
    attribute supportsOptimization : Boolean;
    attribute requiresStatePreparation : Boolean;
  }

  part def ProblemEncoding {
    attribute encodingType : EncodingType;
    attribute requiresLossDistribution : Boolean;
    attribute requiresTailProbability : Boolean;
    attribute requiresObjectiveFunction : Boolean;
  }

  part def BackendProfile {
    attribute backendName : String;
    attribute isPrivate : Boolean;
    attribute maxQubits : Integer;
    attribute maxCircuitDepth : Integer;
    attribute maxShots : Integer;
  }
  part def FeasibilityEvaluation {
    attribute requiredQubits : Integer;
    attribute estimatedCircuitDepth : Integer;
    attribute estimatedShots : Integer;
    attribute estimatedRuntimeHours : Real;
    attribute expectedLossError : Real;
    attribute varError : Real;
  }

  enum def ProblemType {
    enum ProbabilisticRiskEstimation;
    enum CombinatorialOptimization;
    enum EnergySimulation;
    enum LinearAlgebra;
  }
  enum def AlgorithmPattern {
    enum ProbabilityEstimation;
    enum Optimization;
    enum Variational;
    enum Sampling;
  }
  enum def EncodingType {
    enum LossDistributionEncoding;
    enum QUBOEncoding;
    enum HamiltonianEncoding;
    enum LinearSystemEncoding;
  }
  enum def ValidationStatus {
    enum Candidate;
    enum Rejected;
    enum Validated;
  }
}
\end{lstlisting}

The second task formalizes compatibility and dependency constraints directly in the strategy-configuration metamodel. The constraints embedded in \texttt{StrategyInstance} in Listing~\ref{lst:strategyMetamodel} define metamodel-level validity rules rather than application-instance checks: if the structured problem is classified as \texttt{ProbabilisticRiskEstimation}, then the selected algorithm must support probability estimation, and explicitly incompatible encodings such as \texttt{QUBOEncoding} are excluded. Additional metamodel constraints are derived from algorithmic assumptions in the literature, hardware specifications, emerging standards, and quantum software repositories. By embedding these rules directly in the SysML v2 model layer, \qready's implementation can automatically validate configurations and prune invalid combinations early in the modeling process. For the credit-risk example, the problem model created in Section~\ref{subsec:ProblemStructuring} is therefore constrained toward a probability-estimation strategy and away from optimization-oriented encodings before any concrete backend or feasibility values are instantiated.

The third task supports assisted strategy configuration. Building on the structured problem models obtained in the problem structuring phase (Section~\ref{subsec:ProblemStructuring}) and the metamodel-constrained variability structure, \qready performs constrained model completion by identifying admissible configuration paths, proposing candidate strategies aligned with detected structural patterns, and preventing semantically invalid selections. Rule-based reasoning grounded in model semantics provides the primary validation mechanism, while language-model-assisted guidance can complement it when explanatory or exploratory support is useful. This combination separates enforceable semantic constraints from advisory modeling assistance, which is important for reliable and explainable strategy configuration. 

The final task generates a validated strategy instance that conforms to the metamodel in Listing~\ref{lst:strategyMetamodel}. Such strategy instances encapsulate all selected configuration elements under enforced constraints, maintain links to the originating problem model, and record the feasibility values used during validation. They constitute the formal output of this phase and serve as structured input to the hybrid executable workflow modeling phase (Section~\ref{subsec:exeWorkflow}), ensuring methodological continuity across the MBSE pipeline.

\subsubsection{Running Example}
Listing~\ref{lst:creditRiskStrategyInstance} shows a compact instance model for the credit-risk example. The instance links the strategy back to \texttt{CreditRiskProblem\_01}, which was produced by guided problem structuring as a probabilistic risk-estimation problem. More specifically, the listing should be read as the strategy-level continuation of the selected \texttt{F2-SegmentAggregated} reformulation path from Section~\ref{subsec:ProblemStructuring}. That is why the instance binds an iterative amplitude-estimation algorithm, a \texttt{CreditLossDistributionEncoding} corresponding to the bucketed segment-aggregated formulation, a private backend, and feasibility estimates. The strategy is marked \texttt{Validated} at the strategy-configuration level because the evaluation values satisfy the technical QRE constraints: the backend is private, required qubits and circuit depth fit the backend profile, estimated runtime is below four hours, expected-loss error is below 1\%, and VaR error is below 2\%. This status does not by itself imply production approval; governance approval is assessed later in the predictive feasibility phase.

\begin{lstlisting}[language=,caption={Validated strategy instance for the credit-risk running example},label={lst:creditRiskStrategyInstance},basicstyle=\scriptsize\ttfamily,numbers=none,aboveskip=2pt,belowskip=2pt]
package CreditRisk_StrategyInstance {
  public import QREADY_StrategyConfigurationMetamodel::*;

  part creditRiskProblem_01 : ProblemModel {
    attribute problemId = "CreditRiskProblem_01";
    attribute problemType = ProblemType::ProbabilisticRiskEstimation;
    attribute structuringObjective = "Correlated probabilistic simulation plus tail-risk estimation";
  }

  part approvedPrivateBackend_01 : BackendProfile {
    attribute backendName = "Approved private quantum backend";
    attribute isPrivate = true;
    attribute maxQubits = 80;
    attribute maxCircuitDepth = 5000;
    attribute maxShots = 100000;
  }

  part qaeFeasibility_01 : FeasibilityEvaluation {
    attribute requiredQubits = 64;
    attribute estimatedCircuitDepth = 3200;
    attribute estimatedShots = 50000;
    attribute estimatedRuntimeHours = 3.5;
    attribute expectedLossError = 0.008;
    attribute varError = 0.015;
  }

  part creditRiskStrategy_01 : StrategyInstance {
    attribute strategyId = "CreditRiskStrategy_01";
    attribute status = ValidationStatus::Validated;
    ref problem = creditRiskProblem_01;
    part paradigm : HybridQuantumClassical;
    part algorithm : IterativeAmplitudeEstimation;
    part encoding : CreditLossDistributionEncoding;
    part backend = approvedPrivateBackend_01;
    part evaluation = qaeFeasibility_01;
  }
}
\end{lstlisting}


\subsection{Hybrid Executable Workflow Modeling}
\label{subsec:exeWorkflow}
\subsubsection{Research Rationale}
While the strategy configuration and instantiation phase (Section~\ref{subsec:strategyConfigInstantiation}) enables the derivation of formally valid strategy instances, it does not by itself ensure that these configurations can be realized as executable hybrid solutions. In current practice, hybrid quantum–classical applications are typically implemented directly within SDK-specific environments such as Qiskit or D-Wave toolchains, where orchestration logic, backend invocation, iteration control, and data exchange are encoded procedurally~\cite{yeEmuPlatFrameworkAgnosticPlatform2026}. As a result, key aspects such as control-flow semantics, convergence behavior, and hybrid partitioning remain implicit, tightly coupled to specific platforms, and weakly traceable to higher-level design decisions~\cite{wederHybridQuantumApplications2021}.

This disconnect highlights a fundamental limitation: there is no explicit representation that bridges strategy-level configurations and their executable realization. Without such a representation, it is difficult to systematically analyze workflow behavior, reason about alternative execution strategies, or perform feasibility assessment prior to implementation. This gap motivates the need for an explicit workflow modeling layer that captures control flow, component interactions, and data propagation in a structured and analyzable form. By elevating hybrid workflows to first-class modeling artifacts, it becomes possible to ensure consistency between design and execution, while enabling downstream analysis such as predictive feasibility assessment (Section~\ref{subsec:feasibilityAssessment}).

\subsubsection{State of the Art}
As discussed earlier, classical software engineering provides a rich set of modeling notations (e.g., UML, SysML) to support abstraction and analysis. In particular, executable behavior modeling has been extensively studied, supported by well-established notations such as BPMN and state machines, which enable the explicit specification of control flow, data dependencies, and system interactions~\cite{LLM4MDESurvey}. These notations, together with associated execution and analysis techniques, provide a solid foundation for modeling and reasoning about executable workflows in classical systems.

In the quantum software engineering context, workflow-based approaches have become an important direction for orchestrating hybrid quantum--classical applications. QuantME extends BPMN with quantum-specific modeling constructs and transformations for executable quantum workflows~\cite{wederHybridQuantumApplications2021}. Building on this line of work, Beisel et al.~\cite{beisel2025pattern} present a pattern-based approach for generating and adapting quantum workflows from selected quantum computing patterns. Their approach combines a quantum computing pattern language, solution repositories, QuantME modeling constructs, workflow rewriting, packaging, deployment, and execution support, and evaluates the prototype through use cases, runtime measurements, and a user study. This work provides strong evidence that reusable workflow patterns and automation can reduce the complexity of constructing hybrid quantum applications.

The focus of \qready is complementary. Existing workflow-generation approaches primarily start from selected quantum patterns or available implementations and automate their realization as executable workflows. \qready instead places the workflow layer inside a broader MBSE traceability chain, supported by MDE transformations: workflow models are derived from quantum-aware requirements, structured problem models, and validated strategy instances, and then become the basis for predictive feasibility assessment. Thus, QSE workflow automation provides an important foundation, while \qready addresses the upstream and downstream links needed for early, requirements-driven feasibility reasoning.

Simulation plays a central role in MBSE, and MDE techniques are often used to operationalize model execution and analysis for validating design decisions and supporting early-stage feasibility assessment~\cite{incoseINCOSESystemsEngineering2023}. In classical systems, a variety of model execution and simulation techniques have been developed to enable reasoning about workflows, performance, and resource constraints at the system level~\cite{de2022taxonomy, habermehl2022optimization, kinay2024advancing}.
In contrast, such capabilities remain largely absent for hybrid quantum–classical systems. Although simulation is extensively used in quantum computing, existing approaches primarily operate at the level of quantum algorithms and circuits. Current techniques include state vector simulation, which provides exact representations of quantum states but scales poorly with increasing qubit counts; density matrix simulation, which supports modeling noise and decoherence at higher computational cost; and hardware-aware simulation, which incorporates device-specific noise models to approximate real execution environments.
However, these techniques focus on the behavior of isolated quantum components rather than the execution of end-to-end hybrid workflows. As a result, they provide limited support for analyzing system-level properties such as cross-component interactions, resource allocation, and architectural trade-offs, which are essential for feasibility assessment of hybrid quantum–classical applications.

A range of platforms support quantum software modeling and simulation, spanning both hardware-specific and general-purpose environments. For instance, Qiskit is tailored for IBM quantum platforms, while more general tools provide broader modeling and integration capabilities.
The MATLAB Quantum Computing Toolbox\footnote{\url{https://jp.mathworks.com/products/quantum-computing.html}, accessed June 13, 2026.} supports the construction, simulation, and execution of quantum algorithms, with connectivity to external hardware and simulators, leveraging MATLAB’s extensive ecosystem for cross-domain integration. QuTiP~\cite{johansson2012qutip, johansson2013qutip2, qutip_website} focuses on simulating open quantum systems, offering abstractions such as density matrices, noise modeling, and quantum trajectory methods, and integrates with scientific computing libraries in Python. PennyLane~\cite{pennylane} targets hybrid quantum machine learning, enabling integration with classical ML frameworks (e.g., TensorFlow, PyTorch) and providing reusable templates for variational circuits across multiple quantum backends.
While these platforms provide essential capabilities for circuit construction, simulation, and hybrid algorithm development, their abstractions remain largely centered on quantum operations and algorithmic workflows. They offer limited support for modeling executable hybrid workflows at the system level, particularly in representing cross-component interactions, control flow, and design alternatives. Consequently, they are insufficient for enabling systematic design-space exploration and predictive feasibility analysis.

\subsubsection{Research Roadmap}
\qready addresses this limitation through a structured workflow modeling layer that systematically transforms validated strategy instances into executable hybrid workflows. The core objective is to make orchestration logic explicit, analyzable, and reusable. Instead of encoding hybrid coordination in scripts, \qready models sequencing, branching, iteration, synchronization, and quantum-classical invocation boundaries as first-class elements with well-defined semantics. Similarly, recurrent orchestration structures observed in hybrid quantum-classical implementations such as classical outer-loop optimization with quantum evaluation (e.g., VQE/QAOA-style loops) ~\cite{leymannPatternLanguageQuantum2019}, preprocessing followed by quantum core execution~\cite{mancillaPreprocessingPerspectiveQuantum2022}, or sampling-driven iterative refinement as in quantum annealing~\cite{wangTestCaseMinimization2024} can be systematically abstracted into reusable workflow patterns. Formalizing these patterns within SysML v2 enables disciplined composition of heterogeneous computational nodes and early validation of interface compatibility, data dependencies, and partitioning constraints before execution, thereby extending MBSE practice with MDE-style automation for hybrid quantum workflows. 

To achieve the above objective, the first concern is explicit control-flow and execution semantics for hybrid workflows at the modeling layer. These hybrid workflow models capture sequencing, branching, iteration, synchronization, and well-defined invocation boundaries between classical and quantum components. In the credit-risk case, this means making explicit that a classical controller constructs the governed loss model, invokes quantum probability estimation repeatedly inside a VaR search, computes tail metrics, and validates the result against acceptance criteria. Execution configurations, such as precision, confidence, backend selection, and fallback behavior, are therefore modeled as explicit workflow parameters rather than left implicit in SDK scripts.

Similarly, a D-Wave–based workflow may define a boundary where a QUBO formulation is submitted to an annealer, returning sampled solutions and associated energy values to a classical post-processing stage. These invocation boundaries are modeled explicitly to distinguish local classical computation from remote quantum service calls, including data encoding and decoding steps. Further, iterative structures are specified declaratively based on convergence criteria rather than embedded in procedural code. For instance, a quantum annealing workflow may repeat sampling until a stable minimum energy distribution or solution frequency is observed. By modeling such convergence conditions and evaluation feedback explicitly at the workflow level, hybrid orchestration logic becomes transparent, analyzable, and traceable to the strategy instance, instead of being hidden in imperative scripts.

In \qready, hybrid workflows are modeled graphically to support clearer configuration and improved stakeholder comprehension, and SysML v2 behavioral constructs (e.g., actions) are extended with domain-specific notations for quantum execution boundaries, Hamiltonian evaluation, energy estimation, sampling operations, hybrid feedback control, etc. 

Second, interactive and configurable workflow patterns are derived through systematic analysis. Reusable orchestration templates are identified and abstracted from multiple sources: 1) empirical studies and published hybrid algorithm implementations (e.g., VQE, QAOA, quantum annealing workflows), 2) open-source repositories such as Qiskit and D-Wave application examples, 3) SDK documentation and reference architectures, and 4) internally developed experimental prototypes. Through structured mining and analysis of these artifacts, recurring observations (e.g., preprocessing followed by quantum core execution and classical refinement) are formalized as parameterized workflow patterns. These patterns are then configurable according to the selected strategy instance, enabling systematic reuse.

Third, structured composition of classical and quantum computational nodes is supported through targeted extensions of SysML v2. Specific domain notations extend SysML v2 behavioral and structural constructs, enabling explicit representation of classical computation modules, quantum execution units, Hamiltonian evaluation components, sampling blocks, and hardware interaction nodes. Each node type is defined with well-typed interfaces and clearly specified execution roles. Formal composition rules ensure that nodes can only be connected when interface compatibility, data-type alignment, and structural preconditions are satisfied. By extending SysML v2 with these hybrid-specific modeling elements, modularity is preserved and classical–quantum partitioning decisions become explicit, analyzable, and traceable within the workflow model.

Fourth, data dependencies and interface binding are formalized to ensure full executability of the modeled workflows, which is essential for the feasibility analysis. Executable hybrid workflows require precise specification of data flow, parameter propagation, encoding and decoding steps, and intermediate transformation artifacts. \qready therefore introduces explicit modeling constructs to represent inputs, outputs, intermediate results, and binding relationships between classical and quantum components. Type compatibility, dimensional consistency, data format alignment, and strict separation between classical and quantum data domains are validated at the model level prior to execution. By enforcing these constraints, the resulting workflow is not only structurally coherent but also executable in practice, enabling systematic feasibility assessment, performance evaluation, and backend-specific validation. 

\subsubsection{Running Example}
\begin{figure*}[t]
	\centering
	\makebox[\textwidth][c]{\includegraphics[width=\textwidth]{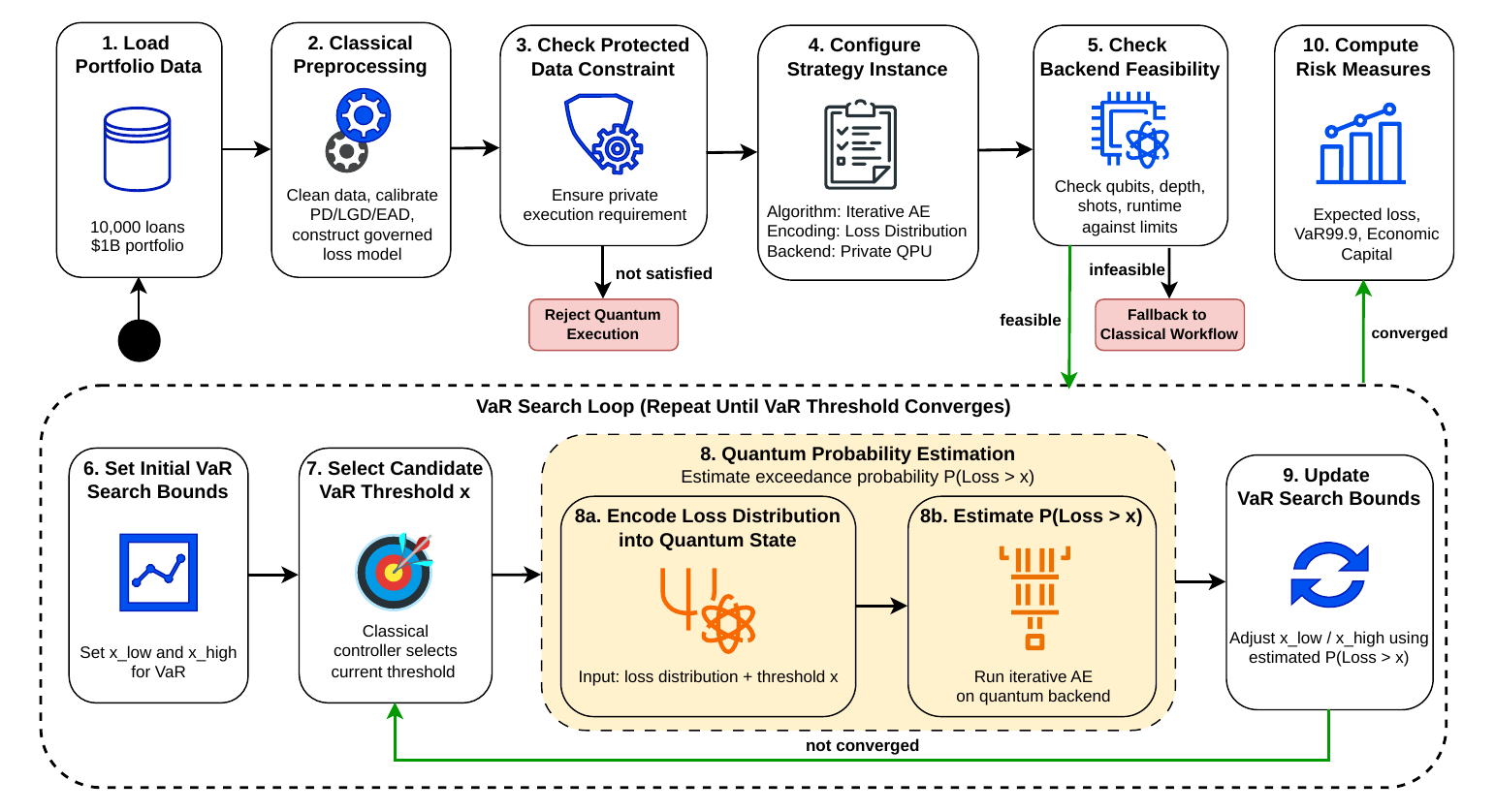}}
		\caption{Illustrative \qready workflow view for the credit-risk running example. This is not a SysML v2 diagram; it is a reader-oriented visualization of the intended executable workflow before the dedicated SysML v2 hybrid workflow notation is implemented. Action boxes show the planned workflow steps; arrows show control, iteration, and evidence flows; and the highlighted quantum-execution region marks where encoded aggregate inputs are submitted to the quantum probability-estimation service. The workflow is aligned with the earlier artifacts: \texttt{CreditRiskProblem\_01} motivates probabilistic risk estimation, \texttt{CreditRiskStrategy\_01} provides the selected strategy configuration, and the QRE constraints reappear as protected-data, backend-feasibility, runtime, accuracy, approval, and evidence-retention checks.}
	\Description{A \qready workflow diagram for credit-risk capital assessment. It organizes the workflow into preparation and configuration, hybrid VaR search execution, evaluation and validation, and governance decision phases. The workflow loads portfolio data, preprocesses the classical loss model, checks protected data constraints, configures the strategy instance, checks backend feasibility, performs quantum probability estimation inside a VaR search loop, computes risk metrics, compares results with the classical benchmark, validates thresholds, records evidence, and routes the result to approval, shadow-mode use, rejection, or fallback.}
	\label{fig:creditRiskWorkflow}
\end{figure*}

The validated strategy instance in Listing~\ref{lst:creditRiskStrategyInstance} is transformed into the planned executable workflow model \texttt{CreditRiskWorkflow\_v1}. This handoff is explicit: the workflow inherits the \texttt{CreditRiskProblem\_01} problem definition, the \texttt{HybridQuantumClassical} paradigm, the \texttt{IterativeAmplitudeEstimation} algorithm choice, the \texttt{CreditLossDistributionEncoding} selected for the bucketed formulation, the \texttt{approvedPrivateBackend\_01} backend profile, and the preliminary feasibility bounds recorded in \texttt{qaeFeasibility\_01}. Figure~\ref{fig:creditRiskWorkflow} is an illustrative workflow view of that model, not a SysML v2 diagram. The dedicated SysML v2-based hybrid workflow notation, including graphical notations for quantum execution boundaries, hybrid loops, and evidence flows, is part of the future implementation of \qready. Even so, the illustrative view is consistent with the strategy instance and earlier artifacts: it realizes the probabilistic risk-estimation problem, embeds the selected probability-estimation strategy into executable control flow, uses the selected loss-distribution encoding, invokes probability-estimation actions on the approved private backend, and carries forward the QRE runtime, backend, data-protection, accuracy, and evidence-retention constraints as explicit workflow checks.

The workflow first performs preparation and configuration: it loads the 10,000-loan portfolio, preprocesses the classical loss model, instantiates the \texttt{CreditLossDistributionEncoding}, enforces the protected-data constraint, binds the validated strategy instance, and checks whether \texttt{approvedPrivateBackend\_01} can satisfy the qubit, depth, shot, and runtime limits already estimated in \texttt{qaeFeasibility\_01}. If these checks fail, the workflow rejects quantum execution or falls back to the classical benchmark. If they pass, the hybrid execution phase realizes the \texttt{IterativeAmplitudeEstimation} choice from Listing~\ref{lst:creditRiskStrategyInstance}: it initializes the VaR search interval and repeatedly invokes quantum probability estimation for tail-probability queries until the VaR loop converges. The evaluation phase computes expected loss, VaR99.9, and economic capital, compares them against the classical benchmark, and validates the 1\% expected-loss error, 2\% VaR error, and four-hour runtime thresholds. The governance phase records evidence for audit and reproducibility and routes the result to model-risk review, governed production use, shadow-mode use, or rejection. Thus, Figure~\ref{fig:creditRiskWorkflow} clarifies how the problem, strategy, backend, and preliminary feasibility information from Listing~\ref{lst:creditRiskStrategyInstance} are realized as first-class workflow behavior in the envisioned SysML v2-based notation.


\subsection{Predictive Feasibility Assessment for Hybrid Quantum Workflows}
\label{subsec:feasibilityAssessment}

\subsubsection{Research Rationale}
Our overall goal is to support the predictive feasibility assessment for hybrid quantum workflows, i.g., determining whether strategy instances and executable workflow models can meet resource, performance, and scalability requirements before committing to physical infrastructure or costly experimentation. 
Early feasibility analysis has been recognized as essential in quantum application research, where hardware access is limited and device characteristics evolve rapidly~\cite{gemeinhardtModelDrivenQuantumSoftware2021, kasiCostPowerFeasibility2023, islamEvaluatingCostEffectiveReconfigurable2025}. Studies on variational algorithms and quantum annealing have shown that performance is highly sensitive to qubit count, connectivity structure, noise levels, and circuit depth, often leading to discrepancies between theoretical promise and practical realizability~\cite{buonaiutoEffectsQuantumHardware2024, pellow-jarmanEffectClassicalOptimizers2024}.

Hybrid computational workflows are particularly sensitive to backend characteristics such as qubit availability, coupling topology, gate duration, and stochastic noise behavior, as well as to classical optimization overhead and iterative execution dynamics reported in the literature on VQE, QAOA, and annealing-based optimization~\cite{wederHybridQuantumApplications2021, pellow-jarmanEffectClassicalOptimizers2024, mccaskeyComposableProgrammingHybrid2021}. 
\qready addresses these challenges through a layered assessment approach that combines backend abstraction, executable workflow simulation, lightweight scaling estimation, and structured feasibility reporting. Rather than developing new simulation engines, \qready leverages established simulation techniques and calibrated noise models through a unified abstraction interface. This approach ensures methodological transparency, comparability across backend scenarios, and informed decision support grounded in documented challenges identified in contemporary quantum computing research.

Feasibility assessment thus serves as the analytical validation layer of \qready. It converts executable hybrid workflows into structured, quantitative feasibility evidence using controlled simulation and lightweight scaling estimation. By separating feasibility assessment from immediate hardware availability while remaining compatible with future empirical benchmarking, we ensure pragmatic evaluation and architectural adaptability. Moreover, the explicit modeling of assumptions, metrics, and constraint evaluations promotes methodological transparency, which supports educational use, knowledge transfer, and clearer communication of hybrid quantum-classical system behavior to both technical and non-expert stakeholders.

\subsubsection{State of the Art}
Recent years have seen increasing attention to hybrid quantum workflows as a means to structure, orchestrate, and optimize quantum–classical applications. Prior work has explored multiple dimensions of this paradigm, including workflow optimization and rewriting to improve execution efficiency~\cite{weder2022analysis}, middleware abstraction for cross-platform deployment and reduced vendor lock-in~\cite{weder2025qunicorn}, and the definition of execution patterns that guide the design and selection of workflow strategies across heterogeneous platforms~\cite{georg2023execution}. In addition, pattern-driven automation has been proposed to generate and adapt workflows~\cite{beisel2025pattern}, as well as techniques to automatically derive workflow models from script-based implementations~\cite{vietz2022splitting}.

Despite these advances, existing approaches largely focus on modeling, generation, and execution optimization, while feasibility analysis remains under-explored. In particular, current works do not systematically account for critical factors such as hardware resource constraints, noise characteristics, and execution uncertainty, nor do they tightly connect workflow design to end-to-end application requirements. 

\subsubsection{Research Roadmap}
The first task defines parametric backend abstraction models. Hardware characteristics such as qubit capacity, connectivity topology, gate durations, coherence times, error rates, and calibrated noise parameters are represented as typed properties of backend definitions. Existing noise models from platforms such as Qiskit are systematically incorporated, embedding device-level noise characterization directly into the backend abstraction rather than treating it as an external simulator setting. 
Parametric constraints capture relationships among these properties (e.g., qubit availability, coupling compatibility, latency and error accumulation bounds), and binding connectors relate backend parameters to workflow metrics such as circuit depth and execution time. Backend scenarios (ranging from idealized simulators to conservative near-term devices) are instantiated as concrete SysML v2 model instances with different property configurations. By separating backend specifications from workflow definitions at the model layer, this approach enables structured, technology-agnostic yet noise-aware feasibility reasoning and comparative analysis across alternative backend configurations. 

The second task performs predictive execution emulation directly on the executable workflow models generated in the hybrid executable workflow modeling phase (Section~\ref{subsec:exeWorkflow}). The workflow engine is executed in a simulation mode, where classical control logic (iteration, branching, optimizer updates) runs as defined, while quantum execution nodes are redirected to appropriate simulators instead of physical hardware. Depending on fidelity and scale requirements, circuit evaluation may use statevector simulation for small problem instances~\cite{guerreschiIntelQuantumSimulator2020}, shot-based noisy simulation (e.g., calibrated noise profiles derived from Qiskit Aer~\cite{anisimovQiskitAerHigh2021}) for realistic device approximation, or annealing simulators for QUBO-based strategies~\cite{yarkoniQuantumAnnealingIndustry2022}. Workflow-level metrics (e.g., circuit depth, gate counts, execution time, sampling variance, and hybrid loop convergence) are collected during emulation and bound to backend abstraction parameters defined in the first task. Doing so enables noise-aware, resource-aware predictive feasibility analysis while preserving a pluggable simulation architecture that allows different simulation engines to be incorporated without modifying the surrounding assessment pipeline.

The third task focuses on deriving a small, well-defined set of resource and performance indicators directly from executable workflow simulations. During predictive execution, the workflow engine records basic metrics such as qubit count required, circuit depth, total gate count, number of circuit evaluations (shots × iterations), and total runtime per hybrid loop. These values are combined with backend parameters (e.g., gate duration and error rate) to estimate approximate execution time and cumulative error likelihood. For larger instances that cannot be fully simulated, simple scaling extrapolations based on observed trends (e.g., linear or quadratic growth with problem size) are applied. The objective is not full performance modeling, but to produce transparent, comparable feasibility indicators that support early decision-making. 

The final task consolidates simulation results and derived metrics into concise predictive feasibility reports. Each report summarizes key indicators such as required qubit capacity, circuit depth, estimated execution time, cumulative error exposure under calibrated noise assumptions, and observed convergence behavior. Identified constraints (e.g., qubit shortages, excessive depth, or instability under noise) are explicitly linked to corresponding strategy or workflow elements, enabling targeted refinement of configuration parameters or structural adjustments. This establishes a clear feedback loop to earlier components, supporting iterative improvement without requiring direct access to experimental hardware.

\subsubsection{Running Example}
\textit{Backend Abstraction.} The predictive assessment starts by instantiating a backend abstraction for \texttt{approvedPrivateBackend\_01}, the backend selected in \texttt{CreditRiskStrategy\_01}. Table~\ref{tab:backend-abstraction} makes the relevant backend assumptions explicit. The numeric limits are consistent with the strategy instance in Listing~\ref{lst:creditRiskStrategyInstance}: private execution is required, the backend provides up to 80 qubits, supports circuit depth up to 5000, and permits up to 100000 shots for the modeled workflow.

\textit{Workflow-Level Execution Simulation.} The executable workflow from Section~\ref{subsec:exeWorkflow} is then evaluated in simulation mode. Consistent with the selected \texttt{F2-SegmentAggregated} reformulation path and the \texttt{CreditLossDistributionEncoding} bound in \texttt{CreditRiskStrategy\_01}, classical actions load and preprocess the portfolio, construct the bucketed loss model, and control the VaR search loop. Quantum execution nodes on \texttt{approvedPrivateBackend\_01} realize the \texttt{IterativeAmplitudeEstimation} choice from Listing~\ref{lst:creditRiskStrategyInstance}: one branch estimates expected loss, while the VaR loop repeatedly evaluates tail probabilities for candidate thresholds until the interval converges. The simulation records the number of calls, shots, circuit depth, runtime, and uncertainty associated with the probability estimates.

\setcounter{table}{2}
\begin{table}[tbp]
\centering
\footnotesize
\caption{Backend abstraction for the running credit-risk example}
\label{tab:backend-abstraction}
\setlength{\tabcolsep}{3pt}
\renewcommand{\arraystretch}{1.04}
\begin{tabularx}{\columnwidth}{@{}>{\raggedright\arraybackslash}p{0.24\linewidth} >{\raggedright\arraybackslash}p{0.20\linewidth} >{\raggedright\arraybackslash}X@{}}
\toprule
\textbf{Backend property} & \textbf{Example value} & \textbf{Role in feasibility assessment} \\
\midrule
Backend type & Approved private backend & Defines the execution environment for probability estimation. \\
Private execution & Yes & Satisfies the protected-data restriction. \\
Maximum qubits & 80 & Caps the encoding size for loss and comparison registers. \\
Maximum circuit depth & 5000 & Caps the transpiled amplitude-estimation circuit depth. \\
Maximum shots & 100000 & Caps repeated probability-estimation calls. \\
Connectivity model & Limited device connectivity & Affects routing overhead and effective depth. \\
Noise model & Gate and readout noise & Supports robustness and error estimation. \\
Gate timing model & Backend-specific duration & Supports quantum execution-time estimation. \\
Queue latency & Estimated / bounded latency & Included in end-to-end runtime prediction. \\
\bottomrule
\end{tabularx}
\end{table}

\textit{Resource and Performance Estimation.} Table~\ref{tab:predictive-feasibility-evaluation} summarizes the resulting predictive evidence. The values are illustrative, but they are interpreted here as workflow-level assessment results that remain consistent with the preliminary estimates recorded earlier in \texttt{qaeFeasibility\_01}: the workflow requires 64 qubits, estimated depth 3200, 50000 shots, 3.5 hours of runtime, 0.8\% expected-loss error, and 1.5\% VaR error. In other words, predictive assessment confirms that the selected \texttt{F2-SegmentAggregated} formulation, algorithm choice, and backend profile still satisfy the technical QRE thresholds at the workflow level, while robustness remains marked as a warning because the modeled noise margin is narrow.

\textit{Feasibility Decision.} The decision rule separates strategy-level validation, workflow-level technical feasibility, and production approval. In the running example, \texttt{CreditRiskStrategy\_01} had already been validated at the configuration stage, and the predictive assessment now confirms that the resulting workflow still satisfies the modeled resource, runtime, privacy, and accuracy constraints under the selected backend assumptions. The marginal robustness result is retained as a technical warning flag, but it does not by itself determine the operational disposition. However, the workflow is not yet approved for production because model-risk governance approval has not been granted. The resulting operational decision is therefore \emph{shadow mode}: the quantum workflow may be retained as a governed research workflow and compared against the classical benchmark, while the classical result remains authoritative for reporting.

\setcounter{table}{3}
\begin{table}[tbp]
\centering
\footnotesize
\caption{Predictive feasibility evaluation for the running credit-risk example}
\label{tab:predictive-feasibility-evaluation}
\setlength{\tabcolsep}{3pt}
\renewcommand{\arraystretch}{1.04}
\begin{tabular}{@{}>{\raggedright\arraybackslash}p{0.27\columnwidth} >{\centering\arraybackslash}p{0.15\columnwidth} >{\centering\arraybackslash}p{0.23\columnwidth} >{\centering\arraybackslash}p{0.16\columnwidth}@{}}
\toprule
\textbf{Feasibility item} & \textbf{Predicted value} & \textbf{Threshold} & \textbf{Result} \\
\midrule
Private backend & Yes & Required & Pass \\
Required qubits & 64 & $\leq 80$ & Pass \\
Estimated circuit depth & 3200 & $\leq 5000$ & Pass \\
Estimated shots & 50000 & $\leq 100000$ & Pass \\
Estimated runtime & 3.5 hours & $\leq 4$ hours & Pass \\
Expected-loss error & 0.8\% & $\leq 1\%$ & Pass \\
VaR$_{99.9}$ error & 1.5\% & $\leq 2\%$ & Pass \\
Noise robustness & Marginal & Acceptable robustness required & Warning \\
Governance approval & Not yet approved & Production approval required & Shadow mode \\
\bottomrule
\end{tabular}
\end{table}


\subsection{Running Example Synthesis}
\label{sec:exampleSynthesis}
The running example illustrates how \qready maintains continuity across requirements, problem structuring, strategy configuration, workflow modeling, and feasibility assessment. Table~\ref{tab:running-example-traceability} makes these connections explicit by showing how each phase contributes one concrete artifact to the next.

\setcounter{table}{4}
\begin{table*}[t]
\centering
\footnotesize
\caption{Cross-phase traceability of the running credit-risk example.}
\label{tab:running-example-traceability}
\begin{tabular}{p{0.15\textwidth}p{0.19\textwidth}p{0.31\textwidth}p{0.25\textwidth}}
\toprule
\textbf{Phase} & \textbf{Primary artifact} & \textbf{Selected running-example content} & \textbf{Carried forward} \\
\midrule
Quantum Requirements Modeling & QRE model & \texttt{projectContext}, \texttt{businessGoal}, operational/resource constraints, success criteria for quarterly credit-portfolio capital assessment & Runtime, privacy, backend, and accuracy obligations that constrain later design choices \\
Guided Problem Structuring & \texttt{CreditRiskProblem\_01} & Probabilistic risk-estimation problem with selected \texttt{F2-SegmentAggregated} bucketed reformulation; \texttt{F1-FacilityLevel} and \texttt{F3-HardwarePilot} retained as alternatives & Problem type, structuring objective, and chosen bucketed loss-distribution path \\
Strategy Configuration \& Instantiation & \texttt{CreditRiskStrategy\_01} & \texttt{HybridQuantumClassical} paradigm, \texttt{IterativeAmplitudeEstimation}, \texttt{CreditLossDistributionEncoding}, \texttt{approvedPrivateBackend\_01}, \texttt{qaeFeasibility\_01} & Selected algorithm, encoding, backend, and preliminary feasibility bounds \\
Hybrid Executable Workflow Modeling & \texttt{CreditRiskWorkflow\_v1} & Workflow view with protected-data checks, backend-feasibility checks, repeated probability-estimation calls in the VaR loop, benchmark comparison, fallback, and evidence retention & Executable control flow and measurable workflow-level behavior for assessment \\
Predictive Feasibility Assessment & Feasibility evidence and governance outcome & Backend abstraction, workflow-level metrics, resource/performance table, robustness warning, shadow-mode decision & Early governed decision on technical feasibility versus production approval \\
\bottomrule
\end{tabular}
\end{table*}

The QRE model first captures the domain context, business goal, operational constraints, resource constraints, and success criteria for the quarterly 10,000-loan, \$1 billion credit-portfolio assessment. Guided problem structuring then turns these requirements into \texttt{CreditRiskProblem\_01}, a probabilistic risk-estimation problem whose objective is correlated probabilistic simulation plus tail-risk estimation, rather than generic optimization. The reformulation step identifies \texttt{F2-SegmentAggregated} as the selected candidate while keeping \texttt{F1-FacilityLevel} and \texttt{F3-HardwarePilot} explicit as alternatives.

Strategy configuration then creates \texttt{CreditRiskStrategy\_01}, which binds the structured problem to a hybrid quantum-classical paradigm, iterative amplitude estimation, \texttt{CreditLossDistributionEncoding}, \texttt{approvedPrivateBackend\_01}, and the feasibility estimates recorded in \texttt{qaeFeasibility\_01}. The workflow step sketches the intended \texttt{CreditRiskWorkflow\_v1} as an illustrative workflow view rather than as a finalized SysML v2 workflow diagram; it shows the behavior that the future hybrid workflow notation must capture, including protected-data handling, backend feasibility checks, repeated VaR-loop probability-estimation calls, benchmark comparison, fallback, and evidence retention. Predictive feasibility assessment then evaluates the workflow against the same QRE thresholds and backend limits.

The resulting decision is deliberately more precise than saying that quantum risk analysis is simply promising. The bucketed hybrid workflow satisfies the modeled hard constraints: private execution, 64 required qubits against an 80-qubit limit, circuit depth 3200 against a 5000-depth limit, 50000 shots against a 100000-shot limit, 3.5 hours against the four-hour runtime target, 0.8\% expected-loss error, and 1.5\% VaR99.9 error. However, the noise robustness margin remains a warning and model-risk governance approval has not yet been granted. The workflow is therefore retained in shadow mode while the classical benchmark remains authoritative for production reporting. Table~\ref{tab:feasibility-decision-outcomes} summarizes the decision outcomes used in this final assessment step. In the running credit-risk example, the shadow-mode path applies: resource, runtime, accuracy, and privacy checks pass, but robustness remains a warning and model-risk governance approval is still pending. This is the kind of early, traceable engineering decision that \qready aims to support before a bank commits governed data, infrastructure, and validation effort.

\setcounter{table}{5}
\begin{table}[tbp]
\centering
\footnotesize
\caption{Feasibility decision outcomes for hybrid quantum-classical workflow assessment}
\Description{Decision matrix for feasibility-assessment outcomes. The matrix lists four outcomes: Approved for production, Shadow mode, Rejected or reconfigure, and Technical warning flag. Each row states the condition under which the outcome applies. The running example follows the Shadow mode outcome while also carrying the Technical warning flag because robustness remains marginal.}
\label{tab:feasibility-decision-outcomes}
\setlength{\tabcolsep}{4pt}
\renewcommand{\arraystretch}{1.08}
\begin{tabularx}{\columnwidth}{@{}>{\raggedright\arraybackslash}p{0.33\linewidth} X@{}}
\toprule
\textbf{Outcome} & \textbf{Decision condition} \\
\midrule
\textbf{Approved for production} & Hard constraints pass, robustness is acceptable, and model-risk governance approval is granted. \\
\textbf{Shadow mode} & Hard constraints pass, but production approval is still pending. This is the running-example outcome. \\
\textbf{Rejected / reconfigure} & At least one hard constraint fails, such as privacy, qubits, depth, shots, runtime, or estimation error. \\
\textbf{Technical warning flag} & Robustness is marginal under the selected backend or noise model and may accompany shadow mode. \\
\bottomrule
\end{tabularx}
\end{table}

\clearpage

\section{Conclusion}
\label{sec:conclusion}

\qready represents a forward-looking vision for engineering hybrid quantum–classical applications through predictive feasibility assessment. Grounded in established MBSE practice and supported by MDE techniques, including mature standards, tools, and best practices, it aims to systematically bridge requirements, system design, and execution considerations within a unified modeling framework. By leveraging proven methodologies from classical software and systems engineering, incorporating emerging knowledge from Quantum Software Engineering (QSE), and integrating quantum computing foundations such as noise models, backend characterization, and quantum simulation, \qready seeks to enable informed decision-making prior to implementation.

The running credit-portfolio example illustrates why this traceability matters. The same application concern is carried from quantum-aware requirements, to a probabilistic risk-estimation problem model, to a validated strategy instance, to an illustrative workflow view, and finally to predictive feasibility evidence. The resulting decision is not a generic claim of quantum advantage, but a governed engineering outcome: the bucketed hybrid workflow is technically feasible under the modeled resource, runtime, privacy, and accuracy constraints, yet remains in shadow mode because robustness is marginal and model-risk approval is pending. This example shows how \qready can support early, auditable decisions before organizations commit sensitive data, infrastructure, and validation effort.

Ultimately, \qready aspires to support the development of robust, scalable, and resource-aware hybrid quantum applications in the NISQ era and beyond.



\bibliographystyle{ACM-Reference-Format} 
\bibliography{Q-Ready.bib}

%


\newpage

\end{document}